\documentstyle[11pt,newpasp,twoside,epsf]{article}
\def\edcomment#1{\iffalse\marginpar{\raggedright\sl#1\/}\else\relax\fi}
\reversemarginpar

\begin{document}
\title{Optical and NIR monitoring of the GRB020405 afterglow}
\author{N. Masetti$^1$, E. Palazzi$^1$, E. Pian$^2$, A. Simoncelli$^3$,
L.K. Hunt$^4$, E. Maiorano$^3$, S. Savaglio$^5$, E. Rol$^6$,
A. Levan$^7$, 
A.J. Castro-Tirado$^8$, A. Fruchter$^9$, J. Greiner$^{10}$, 
J. Hjorth$^{11}$, R. Wijers$^6$ \& E.P.J. van den Heuvel$^6$ 
{\sl on behalf of the GRACE collaboration}}

\affil{$^1$IASF/CNR, Bologna;
$^2$INAF - Astron. Obs. of Trieste;
$^3$Univ. of Bologna;
$^4$IRA/CNR, Florence;
$^5$INAF - Astron. Obs. of Rome and JHU, Baltimore;
$^6$Univ. of Amsterdam;
$^7$Univ. of Leicester;
$^8$IAA, Granada; $^9$STScI, Baltimore;
$^{10}$MPE, Garching; 
$^{11}$Univ. of Copenhagen}

\begin{abstract}
Optical and near-infrared (NIR) observations of GRB020405 started 
about 1 day after the GRB and extended over $\sim$70 days.
Photometry shows that the early decay is consistent with a
single power law of index $\alpha$ = 1.54$\pm$0.06 in all bands.
The late epoch light curves, sampled with HST and VLT, exhibit
a plateau or slight rebrightening around 10-20 days after the GRB. 
This bump can be modeled with a SN2002ap template underlying the afterglow.
Alternatively, the late-epoch data can also be fitted using a power law
with index steeper ($\alpha'$ = 1.85$\pm$0.15) than that of the early
decay phase, in agreement with a late shell collision interpretation.
Spectroscopy indicates that the GRB is at $z$ = 0.691 and that the
host galaxy complex is angularly close to a system of at least two galaxies 
at $z$ = 0.472. $R$-band polarimetry shows that the afterglow is polarized, 
with $P$ = 1.5$\pm$0.4 \% and polarization angle $\theta$ = 
172$^{\circ}$$\pm$8$^{\circ}$.
\end{abstract}

\vspace{-0.5cm}
\section{Introduction}

GRB020405 was detected by the Interplanetary Network on 2002 April 5.02877
UT with a duration of $\sim$40 s (Hurley et al. 2002).
Its optical afterglow was detected by Price et al. (2002a) and confirmed by 
subsequent observations (e.g. Covino et al. 2002).
Optical spectroscopy allowed Masetti et al. (2002a) and Price et al. 
(2002b) to determine the redshift of the GRB, $z$ = 0.691.

Here we report on optical imaging, spectroscopy and polarimetry, and
near-infrared (NIR) imaging of the GRB020405 afterglow acquired as part of
our ongoing programs of optical/NIR follow-up of GRB afterglows at ESO
and other telescopes. We also included the available archival HST
pointings. In particular, our NIR observations allowed the first detection
of this afterglow at these wavelengths. A more detailed presentation will
appear in Masetti et al. (2003, in preparation). 

\begin{figure}
\plotfiddle{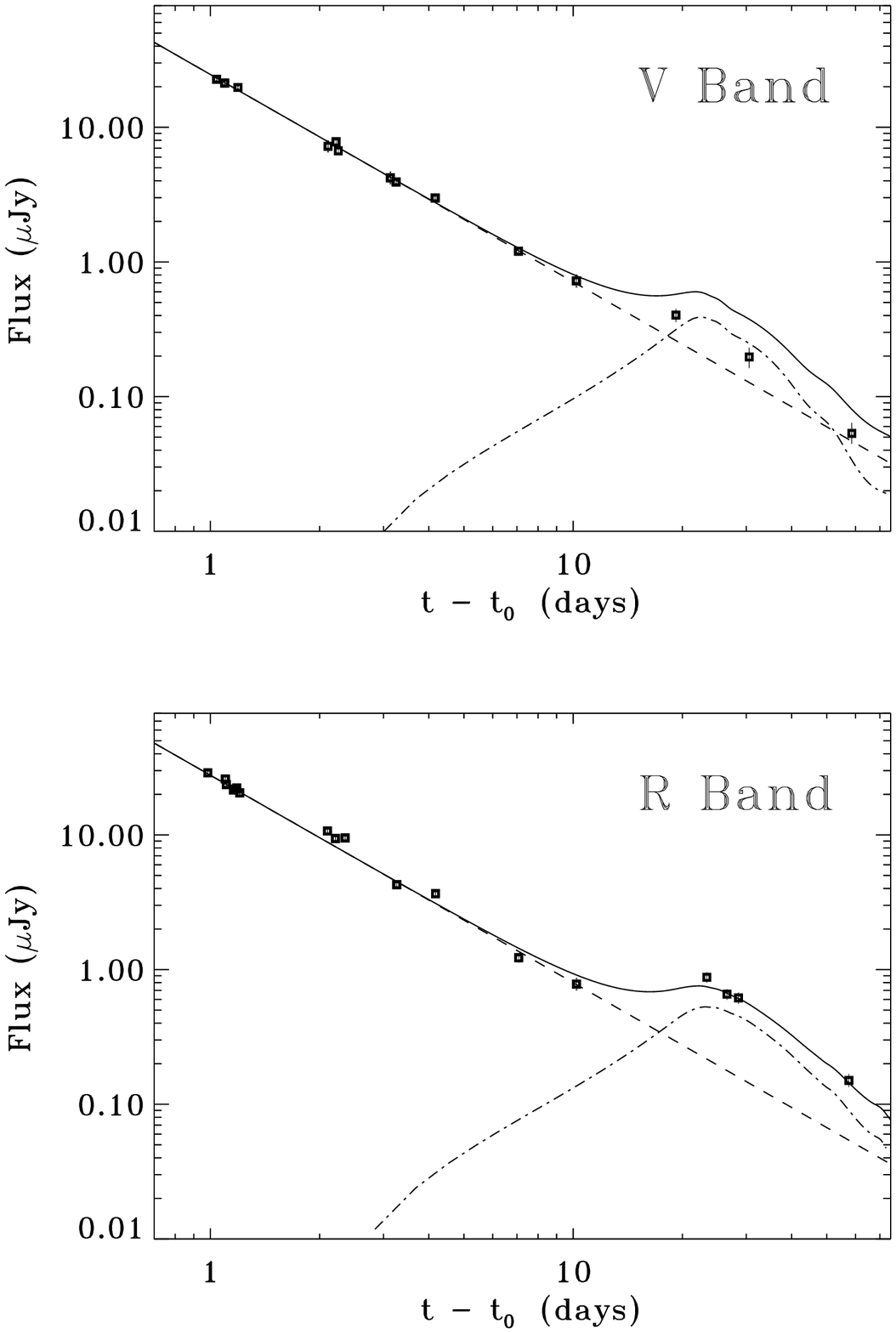}{4cm}{0}{30}{30}{-195}{-95}
\plotfiddle{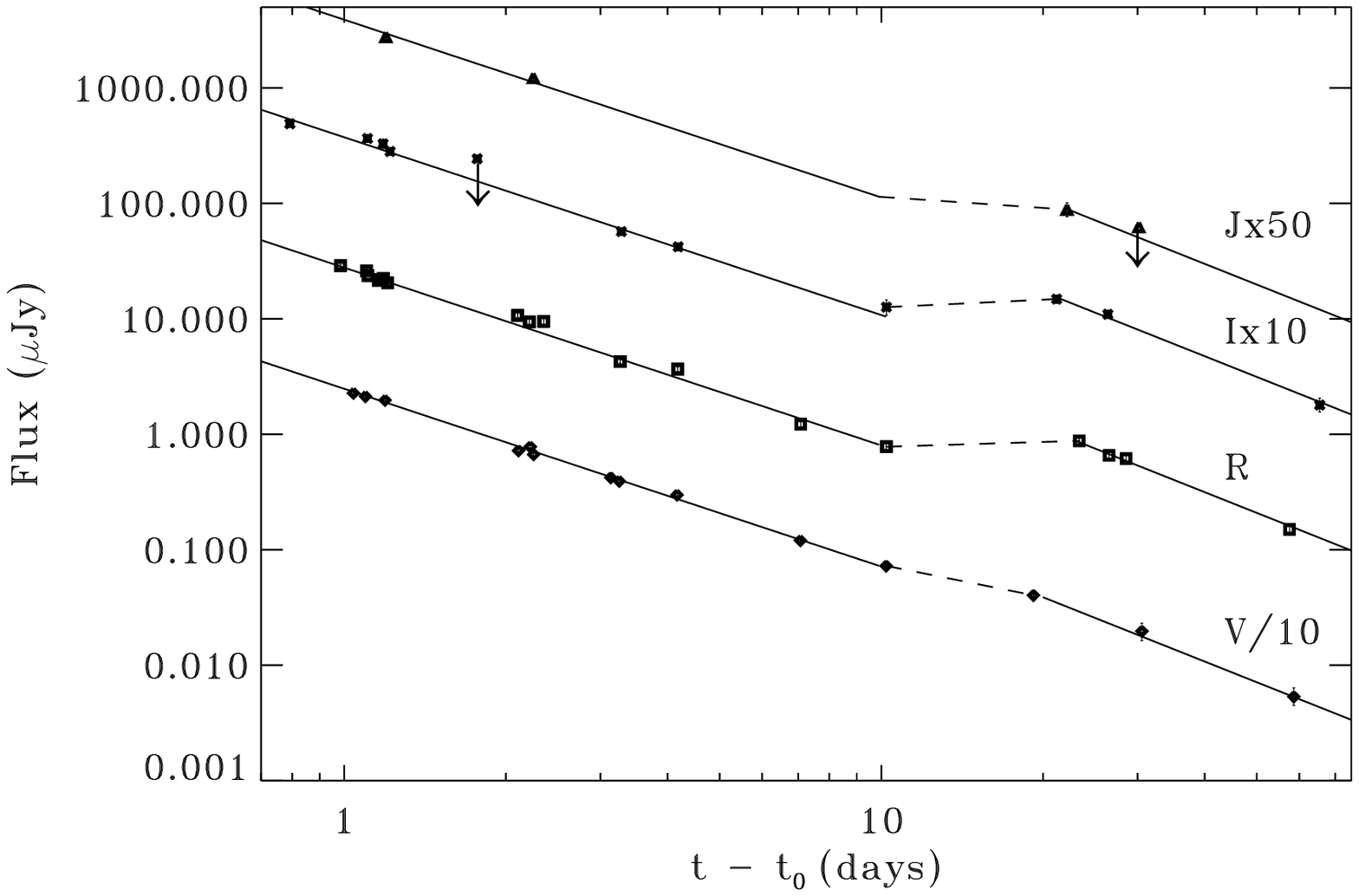}{4cm}{0}{40}{40}{-40}{-65}
\vspace{-2.2cm}
\caption{Host-subtracted light curves of the GRB020405 afterglow, 
corrected for Galactic absorption. {\it Left}: $V$ (upper panel) 
and $R$ (lower panel) data fitted with a power law with $\alpha$ = 1.54 
plus a SN2002ap at $z$ = 0.691, brightened by $\sim$1 mag.
{\it Right}: $VRIJ$ data fitted with a power law with $\alpha$ = 1.54 
(up to day 10) and a power law with $\alpha$ = 1.85 (after day 20). 
The light curves were rescaled in flux for clarity.}
\end{figure}

\section{Observations}

Optical and NIR photometry has been accomplished at several ESO telescopes 
over a period of 10 days, starting on 6 April 2002.
Early optical imaging was acquired also at TNG and WHT (Canary Islands,
Spain) and at the 1.04-m UPSO (Naini Tal, India). The ground-based 
data set has been complemented by archival HST data taken at later epochs.

Optical spectra were taken at VLT-{\it Melipal} on
April 6 and 7 with a dispersion of 5.5 and 2.6 \AA/pix,
respectively. In the second spectroscopic observation the slit was rotated
of about 40$^{\circ}$ with respect to the North-South direction in order to
include in the slit the OT and two nearby galaxies (see Sect. 3).
Furthermore, a $R$-band polarimetric measurement was acquired at VLT-{\it
Melipal} on April 6.

\section{Results}

{\sl Light curves.}
In order to take into account the contribution of the host galaxy to the 
optical transient (OT) emission in the ground-based photometry, we followed 
two ways: (i) we fitted the light curves with a simple power law
plus constant; (ii) we measured, on the latest available HST images, 
the contribution of the host galaxy complex within an aperture radius 
matching the ground-based telescopes PSFs.
The two methods gave consistent results, and showed that the host
contribution to the OT luminosity is of the order of few percent 
(and thus negligible within the uncertainties) in the first days of
observation.

\begin{figure}
\plotfiddle{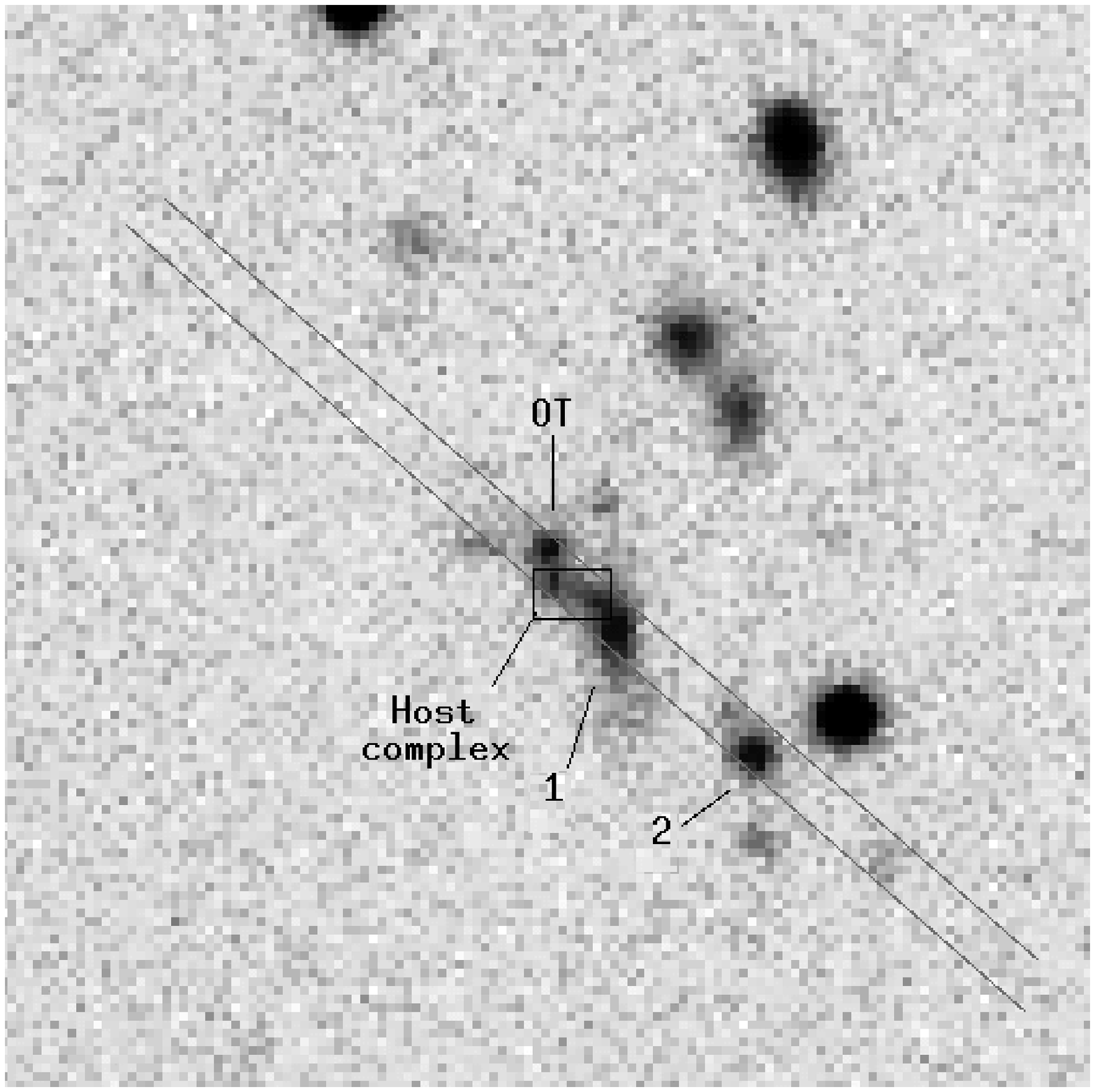}{4cm}{0}{30}{30}{-195}{-80}
\plotfiddle{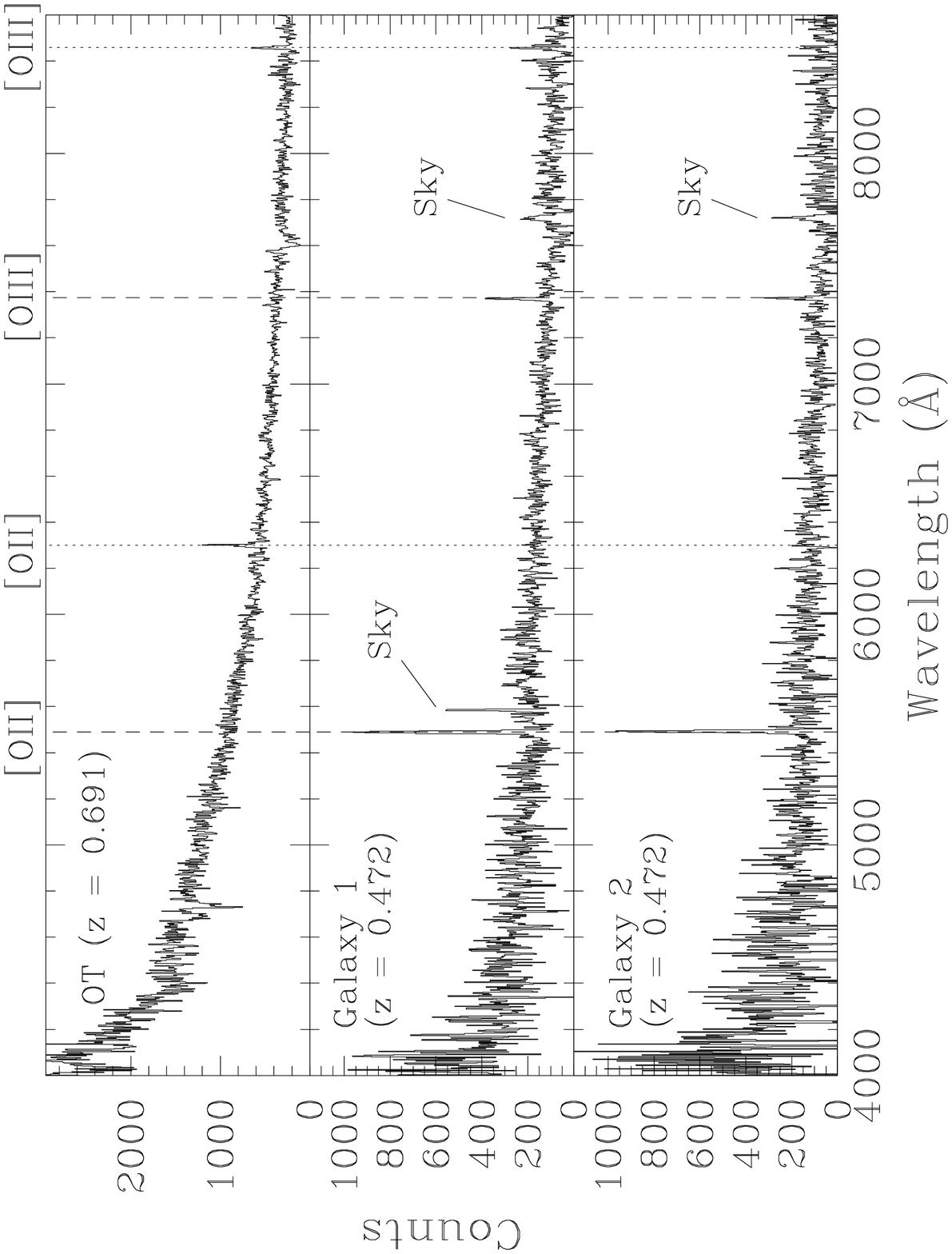}{4cm}{-90}{30}{30}{-30}{265}
\vspace{-3.7cm}
\caption{{\it Left}: VLT-FORS1 $R$-band image, acquired on 15 April 2002,
of the GRB020405 host galaxy complex. The diagonal lines indicate the slit
position of the VLT spectrum taken on April 7. North is at top, East to 
the left; the field is about 25$''$$\times$25$''$. {\it Right}: spectra 
of OT, Galaxy 1 and Galaxy 2, as marked on the VLT image. 
[OII] and [OIII] emission lines are detected at $z$ = 0.691 in the OT 
spectrum (short dash) and at $z$ = 0.472 in the spectra of the two 
closeby galaxies (long dash).} 
\end{figure}

The host-subtracted ground-based optical measurements are fitted by a single
power law decay, with index $\alpha$ = 1.54$\pm$0.06. 
The host contribution in the HST images was estimated and subtracted 
as well by applying a PSF-based method for the OT removal. The addition of 
the HST data points to the $VRI$ light curves made it apparent the presence 
of a deviation from the early single power-law behaviour, i.e. a slight 
rebrightening followed by a decay (Fig. 1, left). In the $RIJ$ light curves, 
this ``bump" is equally well fitted with an emerging SN akin 2002ap, and
1.3 mag brighter, at the OT redshift ($z$ = 0.691), and with
a SN1998bw dimmed by 0.6 mag (see also Price et al. 2002b). $V$-band data
(Fig. 1, upper left) are instead poorly fitted by either SN.
Alternatively, the bump may be interpreted with a shell collision
re-energization scenario, as proposed by Kumar \& Piran (2000) and 
Beloborodov (2002). 
This explanation would be supported by the satisfactory fit of the light 
curves past the bump (from day 20 to 70 after GRB) with a second power 
law of index $\alpha'$ = 1.85$\pm$0.15 (Fig. 1, right).

\medskip
\noindent
{\sl Spectra.}
The spectra acquired on April 6 and 7 with VLT-FORS1 show Balmer, [OII] and
[OIII] lines in emission and FeII and MgII in absortion.
From these, we measure a redshift $z$ = 0.691$\pm$0.002 for the GRB.
Both absorption and emission features in the OT spectrum are consistent with
this value. 
To acquire both the spectrum of the OT and those of Galaxies 1 and 2
located at 2$''$ and $\sim$6$''$ southwest of it, on April 7 the
1$''$-wide FORS1 slit was rotated by 40$^\circ$ towards East with respect
to the N-S direction (Fig. 2, left panel).

From the detection of [OII] and [OIII] emission lines in their spectra,
both Galaxies 1 and 2 appear to be at a substantially lower redshift, 
$z$ = 0.472$\pm$0.002, than the OT host galaxy (Fig. 2, right). Thus, 
although angularly close to the host galaxy complex of GRB020405, 
Galaxy 1 is not interacting with the GRB host galaxy, as formerly proposed
by Masetti et al. (2002b). The preliminary estimate of the redshift of 
Galaxy 1 was incorrect, due to an improper subtraction of the host complex 
contribution from its spectrum.

\medskip
\noindent
{\sl Optical-NIR spectral flux distribution.}
Using photometry data, we have constructed 5 optical-NIR 
broadband spectra. 
The data points were corrected for the Galactic absorption 
($E(B-V)$ = 0.055; Schlegel et al. 1998; Cardelli et al. 1989) and 
converted into fluxes according to Fukugita et al. (1995) for the 
optical and to Bersanelli et al. (1991) for the NIR. 
The spectra of the first two epochs were not corrected for the host
contribution because this is known for the $VRIJ$ bands only; however,
given this was quite modest at those epochs, we simply added a 
5\% error in quadrature to the uncertainties in the optical-NIR fluxes.

The two earliest spectra clearly show a break
at 2.5$\times$10$^{14}$ Hz, i.e. in the $J$ band, with slopes
$\beta_{\rm NIR}$ = 0.65$\pm$0.2 and $\beta_{\rm opt}$ =
1.3$\pm$0.2.
This break can be interpreted as the synchrotron cooling frequency
$\nu_{\rm c}$ in the simplest case of a spherical fireball expanding in a
homogeneous medium (Sari et al. 1998) and with electron distribution index
$p \sim$ 2.6.
Broadband spectra of the three following epochs, made with $VRI$ points, 
were instead plotted by subtracting the host contribution. 
The spectral slopes on these three epochs are consistent with the 
optical one on the first two epochs.
The broad-band spectra of the OT emission during and after the 
``bump" have a steeper power law shape, $\beta$ = 3.5$\pm$0.5, 
in agreement with the findings of Price et al. (2002b).

\medskip
\noindent
{\it Polarimetry.}
Our $R$-band polarimetry data indicate for the 
OT a linear polarization $P_{\rm OT}$ = 1.5$\pm$0.4 \% 
and a polarization angle $\theta_{\rm OT}$ =
172$^{\circ}$$\pm$8$^{\circ}$. This result is corrected for possible
instrumental and interstellar polarization by using field stars and 
polarization standard stars.
This value for the polarization is consistent with those measured
by Covino et al. (2002), but is at variance with that ($P \sim$ 10\%)
obtained by Bersier et al. (2002) from $V$-band observations
acquired nearly simultaneously with ours.


\begin{references}

\reference
Beloborodov, A.M. 2002, ApJ, submitted {\tt (astro-ph/0209228)}
\reference
Bersanelli, M., Bouchet, P., \& Falomo, R. 1991, A\&A, 252, 854
\reference
Bersier, D., et al. 2002, ApJ, submitted {\tt (astro-ph/0206465)}
\reference
Cardelli, J.A., Clayton, G.C. \& Mathis, J.S., 1989, ApJ, 345, 245
\reference
Covino, S., et al. 2002, A\&A in press {\tt (astro-ph/0211245)}
\reference
Fukugita, M., Shimasaku, K., \& Ichikawa, T. 1995, PASP, 107, 945
\reference
Hurley, K., Cline, T., Frontera, F., et al. 2002, GCN \#1329
\reference
Kumar, P., \& Piran, T. 2000, ApJ, 532, 286
\reference
Masetti, N., Palazzi, E., Pian, E., et al. 2002a, GCN \#1330
\reference
Masetti, N., Palazzi, E., Maiorano, E., et al. 2002b, GCN \#1375
\reference
Price, P.A., et al. 2002a, GCN \#1326
\reference
Price, P.A., et al. 2002b, ApJ, submitted {\tt (astro-ph/0208008)}
\reference
Sari, R., Piran, T., \& Narayan, R. 1998, ApJ, 497, L17
\reference
Schlegel, D.J., Finkbeiner, D.P., \& Davis, M. 1998, ApJ, 500, 525

\end{references}
\end{document}